\documentclass[lettersize,journal]{IEEEtran}

\usepackage{cite}
\usepackage{pgfplots,tikz,subcaption,varwidth,fixltx2e,paralist,graphicx,algorithmic,amsmath,amssymb,amsfonts}
\usepackage{algorithmic}
\usepackage{graphicx}
\usepackage{subcaption}
\usepackage{mwe}
\RequirePackage{filecontents}
\usepackage{textcomp}
\usepackage{xcolor}
\usepackage{todonotes}
\usepackage{float}
\usepackage{url}
\usepackage{bbm}
\usepackage{lipsum,lineno}
\usepackage{amsthm}
\usepackage[normalem]{ulem}
\usepackage{siunitx}
\usepackage{dsfont}
\usepackage{scalerel}
\usepackage{tikz}
\usetikzlibrary{svg.path}
\usetikzlibrary{backgrounds}
\def\BibTeX{{\rm B\kern-.05em{\sc i\kern-.025em b}\kern-.08em
		T\kern-.1667em\lower.7ex\hbox{E}\kern-.125emX}}

\definecolor{orcidlogocol}{HTML}{A6CE39}
\tikzset{
	orcidlogo/.pic={
		\fill[orcidlogocol] svg{M256,128c0,70.7-57.3,128-128,128C57.3,256,0,198.7,0,128C0,57.3,57.3,0,128,0C198.7,0,256,57.3,256,128z};
		\fill[white] svg{M86.3,186.2H70.9V79.1h15.4v48.4V186.2z}
		svg{M108.9,79.1h41.6c39.6,0,57,28.3,57,53.6c0,27.5-21.5,53.6-56.8,53.6h-41.8V79.1z M124.3,172.4h24.5c34.9,0,42.9-26.5,42.9-39.7c0-21.5-13.7-39.7-43.7-39.7h-23.7V172.4z}
		svg{M88.7,56.8c0,5.5-4.5,10.1-10.1,10.1c-5.6,0-10.1-4.6-10.1-10.1c0-5.6,4.5-10.1,10.1-10.1C84.2,46.7,88.7,51.3,88.7,56.8z};
	}
}

\newcommand\orcidicon[1]{\href{https://orcid.org/#1}{\mbox{\scalerel*{\begin{tikzpicture}[yscale=-1,transform shape]\pic{orcidlogo};\end{tikzpicture}}{|}}}}

\usepackage[hidelinks]{hyperref}
\definecolor{internationalkleinblue}{rgb}{0.0, 0.18, 0.65}
\definecolor{aa}{rgb}{0.06, 0.2, 0.65}
\definecolor{green}{rgb}{0.0, 0.2, 0.13}
\definecolor{red}{rgb}{0.79, 0.0, 0.09}
\definecolor{indigo}{rgb}{0.0, 0.25, 0.42}

\begin{document}
	
	\title{Effects of Small-Scale User Mobility on Highly Directional XR Communications}

	\author{\IEEEauthorblockN{ Asad Ali$^{1,2}\orcidicon{0000-0001-6293-4771}$, Olga Galinina$^{1}$\orcidicon{0000-0002-5386-1061}, {Jiri Hosek$^{2}$\orcidicon{0000-0002-8382-9185}}, and {Sergey Andreev$^{1,2}$\orcidicon{0000-0001-8223-3665}}} \\
		\IEEEauthorblockA{
			$^{1}$Tampere University, Tampere, Finland \\
			$^{2}$Brno University of Technology, Brno, Czech Republic\\
			\{asad.ali, olga.galinina, sergey.andreev\}@tuni.fi, hosek@vut.cz} 
		\vspace{-5mm}
	}
	
	\maketitle
	
	\thispagestyle{plain}
	\pagestyle{plain}

	\begin{abstract}
		The development of next-generation communication systems promises to enable extended reality (XR) applications, such as XR gaming with ultra-realistic content and human-grade sensory feedback. These demanding applications impose stringent performance requirements on the underlying wireless communication infrastructure. To meet the expected Quality of Experience (QoE) for XR applications, high-capacity connections are necessary, which can be achieved by using millimeter-wave (mmWave) frequency bands and employing highly directional beams. However, these narrow beams are susceptible to even minor misalignments caused by small-scale user mobility, such as changes in the orientation of the XR head-mounted device~(HMD) or minor shifts in user body position. This article explores the impact of small-scale user mobility on mmWave connectivity for XR and reviews approaches to resolve the challenges arising due to small-scale mobility. To deepen our understanding of small-scale mobility during XR usage, we prepared a dataset of user mobility during XR gaming. We use this dataset to study the effects of user mobility on highly directional communication, identifying specific aspects of user mobility that significantly affect the performance of narrow-beam wireless communication systems. Our results confirm the substantial influence of small-scale mobility on beam misalignment, highlighting the need for enhanced mechanisms to effectively manage the consequences of small-scale mobility.
	\end{abstract}

	\vspace{-5mm}
	\section{Introduction}	
	Extended reality (XR) headsets are envisioned to be as lightweight and compact as standard eyeglasses. The compact design introduces hardware constraints such as limited processing power, memory, and battery capacity. To meet the expected Quality of Experience (QoE) for XR applications, XR headsets might need to offload computation tasks to the network. This calls for wireless connectivity that offers exceptional reliability, ultra-high data rates, and low latency. By utilizing millimeter-wave (mmWave) frequency bands, 5G and beyond networks open up possibilities for XR due to their immense available bandwidth~\cite{wang2023road}. However, wireless communication at these frequencies requires effective beam management, which involves identifying and maintaining the best beam pairs for communication~\cite{heng2021six}. Unfortunately, these highly directional beam pairs are prone to misalignment when the user equipment (UE) is moving.

	Traditionally, research on wireless communication systems predominantly focused on large-scale mobility, where the movement of UE is based on the general trajectory of the user. This form of mobility has been extensively studied within the context of wireless communication systems, often overlooking small-scale mobility, which refers to the lateral shifts and rotations of the UE that occur closer to the main trajectory of the user, or movement experienced by the UE when the user is within a confined space. Unlike large-scale mobility, which involves significant changes in the UE's location such as another geographical location, small-scale mobility encompasses comparatively smaller movements due to posture adjustments or head turns.
	
	For XR applications that result in high user mobility, such as gaming, the frequent movement of UE leads to rapid and abrupt changes in the wireless channel negatively impacting link quality. The challenge is further exacerbated by blockages, deafness, and propagation loss from factors like object and human body shadowing. In the context of XR, antenna arrays integrated into head-mounted devices~(HMDs) worn by users form highly directional beams directed at the access node. These beams change direction as the user turns their head. Even minor misalignments can result in significant link degradation or outages~\cite{khayrov2023packet}, primarily due to the delay in beam realignment. This highlights the importance of considering small-scale user mobility in scenarios involving highly directional communication. To gain deeper insights into the effects of small-scale mobility, it is crucial to observe and analyze real user mobility data. Leveraging real-world data provides a more comprehensive understanding of the aspects of mobility that significantly affect wireless communication such as movement pattern, type of movement, and speed \mbox{of movement.}
	
	In~\cite{stepanov2021statistical}, the authors perform statistical analysis and modeling by considering micro-mobility (i.e., more subtle movements as compared to small-scale mobility) of a user standing still under different applications. This is achieved by adopting a direct measurements methodology involving the usage of a laser attached to the UE and a camera. In~\cite{grannemann2020urban}, the authors investigate the impact of using phased antenna arrays for UE under small-scale mobility and compare the results to a reference horn antenna measurement. The authors used angle of arrival (AoA) and angle of departure (AoD) measurements to assess the achievable data rates, directional link opportunities, and robustness. The results show considerable losses in data rate due to small beam misalignments with significant and irregular variations. The study in ~\cite{ichkov2021millimeter}, explores mmWave beam misalignment effects on small- and large-scale user mobility with the help of channel and mobility measurements in an outdoor setting. The beam misalignment due to small-scale mobility strongly depends on the mobile device use case.
	
	The main objective of this article is to investigate the impact of small-scale user mobility on wireless communication in XR applications. In what follows, we provide an overview of the challenges posed by small-scale mobility during wireless communication with highly directional beams in the mmWave frequency band, specifically for XR applications. We also outline potential approaches to resolving the challenges associated with small-scale mobility. Further, to advance our understanding of small-scale mobility and its implications on highly directional communication, we collect and present a dataset of user mobility during VR gaming sessions. This dataset is subsequently employed to evaluate the effects of user mobility on highly directional communication during XR usage and inform on the aspects of mobility that negatively impact wireless communication.

	\vspace{-3mm}
	\section{Small-Scale Mobility and Highly Directional Communication}
	
	\subsection{Small-Scale Mobility in XR}
	
	Throughout an immersive XR session, the UE experiences continuous small-scale mobility due to factors such as user behavior, content attributes, and interactions with the physical and digital environment. For example, a seated XR user viewing 360-degree videos tends to exhibit a higher degree of rotational movements~\cite{magnani2020head}. On the other hand, a walking user typically performs synchronized cyclic head movements with each step~\cite{hicheur2005head}, supplemented by rotational movements to explore their surroundings.
	
	In a fully mobile XR HMD setup, various sensory data are transmitted to a remote XR server. These data include information about user head motion, eye tracking, simultaneous localization and mapping (SLAM), light detection and ranging (LiDAR), and six degrees of freedom (6DoF) sensors, as well as user input. The server renders and encodes frames based on the reported motion data and sends them back to the XR device to be decoded and shown on the XR display panel. As the XR user moves in the real world, whether laterally or by turning their head, their field of view adapts to mirror these changes in the digital world. This real-time adjustment ensures that the displayed digital content aligns with the user's perspective thereby preserving the immersive experience.
	
	As the user interacts with the digital world, the position and orientation of the antennas on the XR HMD change relative to the access node. Depending on the user's behavior and reaction to the digital content, the small-scale mobility experienced by the UE can be slow and continuous or sudden and abrupt. As the UE's antenna array position and orientation change, the beams adapt to maintain seamless communication with the access node as the user's perspective alters in response to the dynamic digital content.
	
	As users turn their heads to navigate within immersive XR environments, small-scale mobility manifests in faster and more abrupt changes in orientation as compared to lateral movement~\cite{magnani2020head}. These sudden and abrupt movements initiated by the user to alter their perspective during scene transitions can potentially influence their reactions. The high capacity and reliable connectivity facilitated by highly directional beam is crucial for the closed-loop delay-sensitive interaction between user actions (input and sensor data) and the rendered frames displayed to the user.

	\begin{figure}[!t]
		\centering 
		\includegraphics[width=0.49\textwidth, angle=0]{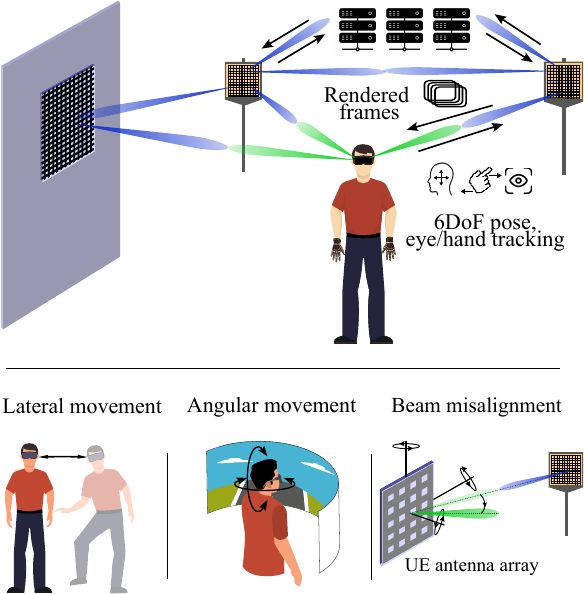}
		\caption{Illustration of directional XR communication and beam misalignment due to small-scale mobility.}\label{fig:dep}
	\end{figure}

	\subsection{Wireless Challenges due to Small-Scale XR Mobility}	
	From a wireless communications perspective in XR applications, it is crucial to maintain synchronization between the render rate and the display rate. This helps prevent the effects such as frame judder, which can result in issues including double images, ghosting, and blurring. Mismatches between the displayed frame and the sensory expectations degrade the QoE of an XR session. Therefore, minimizing frame judder is essential for delivering a seamless and immersive XR experience, one that minimizes the likelihood of \mbox{motion sickness.}
	
	In scenarios with high mobility, the performance of the connection formed by highly directional beams is adversely affected by changes in the environment between the transmit and the receive antennas. The AoA and AoD change rapidly, requiring frequent beam alignment, which incurs significant overhead. Highly directional beams are more susceptible to degradation or disconnection due to even minor misalignments resulting in significant signal-to-noise ratio (SNR) degradation and causing short-term outages~\cite{liu2017millimeter}. 
	
	In the context of mobility, short-term outages refer to temporary interruptions in wireless connection that occur when beams become misaligned and last until beam management procedures realign the beams. In high-mobility scenarios, the time provisioned for a beam to realign based on the channel conditions during the initial misalignment may not be sufficient. Even after switching beams, the newly selected beam pairs can become misaligned due to the rapid movement \mbox{of the device.}
	
	Small-scale mobility introduces beam misalignment through both lateral and angular movements, which alters AoA and AoD at both the access node and the \mbox{UE sides.} However, the most pronounced changes occur at the UE side. Slight lateral movements during communication have the risk to result in outages due to signal obstruction or stepping out of the coverage area, requiring the selection of a new beam. Consequently, by causing momentary connectivity loss, these outages disrupt the immersive perception for the user and greatly degrade the QoE. 
	
	Furthermore, the speed of the user's head movement directly impacts the data size transmitted from the access node to the UE \cite{kim2020motion} to ensure a smooth and immersive experience. As the user's head motion prompts changes in the displayed view, bursts of new frames are needed to accommodate these changes. Consequently, greater user mobility results in increased data transmission, while at the same time, higher mobility also demands more capacity and reliability. This disparity between the user's movements and the link performance in terms of reliability and capacity highlights the need for efficient handling of user mobility, especially \mbox{small-scale mobility.}
	
	The small-scale mobility experienced by UEs is further aggravated by blockers, making it even more challenging to maintain reliable communication links. The highly directional beams in the mmWave frequency bands are susceptible to disruptions caused by blockages. Primarily, three types of blockages commonly need to be considered: (i) static blockers, such as buildings and trees; (ii) dynamic blockers, including neighboring users; and (iii) self-blockages caused by user body parts like the head or arms. While static blockages are effectively modeled and not explicitly considered in indoor mmWave networks, accurately predicting dynamic and self-blockages relies on understanding human behavior, which varies depending on the environment.
	
	Employing real-world mobility data allows for a more accurate assessment of the challenges posed by small-scale mobility and aids in the identification of specific mobility aspects that significantly influence wireless communication performance. The intricate nature of small-scale mobility introduces complexities that disrupt the connections between access nodes and XR formed by highly directional beams. Small-scale mobility experienced by UE during XR sessions is considerably different from the typical large-scale mobility and requires special consideration grounded in realistic scenarios. Unlike large-scale mobility, which involves more gradual and predictable movements, small-scale mobility can lead to frequent and less expected beam misalignments and disruptions in signal transmission. For this reason, it is essential to consider real mobility data during an XR session to quantify the impact on wireless communication, notably for highly \mbox{directional beams. }
	
	\vspace{-3mm}
	\subsection{Addressing Challenges of Small-Scale Mobility}	
	To ensure seamless and immersive XR experiences, side-information and learning-driven beamforming can be employed to mitigate the impact of small-scale mobility during highly directional communication for XR. Beam management is a central aspect in ensuring seamless and immersive XR experience, particularly in addressing the challenges posed by small-scale mobility. Efficient beam switching in response to user movement is essential to counter the performance degradation caused by such mobility. 
	
	XR HMDs are equipped with an inertial measurement unit (IMU) that enables real-time tracking of HMD orientation, including the antenna array. Leveraging IMU data improves beam management, resulting in faster and more agile beam switching. By predicting the AoA and AoD using IMU data, XR communication systems should be able to reduce communication outages caused by abrupt changes in user orientation. This integration of motion sensor data enhances the responsiveness of XR communication systems, hence providing users with a smoother and more immersive experience. Statistical analysis of user movements based on actual XR device data can offer valuable insights into user mobility and help develop intuition to improve XR beamforming. 
	
	Machine learning (ML) algorithms can analyze user's historical movement patterns and real-time sensor data from the headset's IMU. ML algorithms may also be used to predict  user's future movement trajectory and anticipate potential changes in the AoA and AoD of the communication beams. As the user starts to turn their head or move, a suitable ML algorithm can proactively suggest the preferred beam direction to maintain a stable link with the access node. By employing ML-aided beam switching at both the UE and the access node, XR devices may dynamically adapt and respond to user movements with faster beam selection. This translates into improved reliability and capacity of the wireless link formed by the highly directional beam pair and the XR QoE.

	Another opportunity is using an ML-based approach for on-demand beam alignment (in contrast to a periodic beam alignment case), where the beam alignment procedure runs every time when the UE experiences an outage. With an ML-based approach, the UE can rely on IMU and historical channel data to anticipate an outage before it occurs and proactively perform beam realignment.

	In summary, the aforementioned approaches offer promising solutions to address the challenges arising from small-scale mobility in XR communications. In the following section, we present our numerical results derived from real-world mobility data collected during XR gaming sessions, which illustrate the challenges stemming from user small-scale mobility. Additionally, we identify specific aspects of small-scale mobility that impact highly directional communication during XR \mbox{gaming scenarios.}

	\section{Evaluation of Small-Scale Mobility Effects on XR Communications}	
	In this section, we experimentally evaluate the impact of user small-scale mobility on highly directional communication by using mobility data collected from participants during VR gaming sessions. We further describe the user mobility dataset and simulation setup, present our selected numerical results, and analyze the impact of user mobility. This study aims to identify and quantify the specific aspects of small-scale mobility that significantly affect highly directional communication for XR applications, with a particular focus on VR gaming.
	
	\subsection{User Mobility Dataset}
	We use real-world mobility data obtained during VR gaming sessions to assess the impact of small-scale mobility on the performance of highly directional beams. Our approach involves gathering user mobility data from VR gaming scenarios, providing valuable practical insights into small-scale mobility in XR, which are then utilized to evaluate its impact on wireless performance.
	
	We collected mobility data from 40 participants engaged in VR gameplay using an HTC Vive Pro headset. This headset was equipped with 6DoF IMU sensors to accurately track body and head movements. Additionally, user movement was captured by two infrared (IR) devices mounted on tripod stands. The combined data from the headset's IMU sensors and the IR stations provided precise position and orientation information. The VR headset was connected to a high-performance PC workstation featuring an Intel Xeon 6-core CPU, 16 GB of RAM, and a GeForce GTX 1650 GPU. Using OpenTrack, an open-source head-tracking software~\cite{Opentrack} running on the workstation, we recorded the headset's position and orientation. The participants enjoyed a 5m × 5m play area. These measurements were obtained indoors while participants played a game and allowed for real-time collection of orientation and position data. 
	
	The data was collected for 5 minutes from each participant, with the mean sampling rate of approximately 250 Hz. The data was stored in .csv file format, containing 7 fields: 
	recording time, displacement of the HMD position in meters (x, y, z), and angular displacement of the HMD (yaw, pitch, roll).
	The yaw and roll angles ranged from -180 to 180 degrees, while the pitch angle ranged from -90 to 90 degrees. To foster broader community engagement and facilitate further research and benchmarking, this user mobility dataset was made publicly available as part of this publication~\cite{myData}.
	
	\subsection{Evaluation Methodology and Results}	
	To investigate the impact of user mobility on directional communication, we conducted additional simulations that focused on the interactions between an access node and a wireless VR headset. These simulations employed realistic antenna movement patterns based on the collected dataset we described earlier.  The simulations were carried out within a 3D environment, where both the access node and the VR headset were equipped with antenna arrays and adhered to IEEE 802.11ad/ay communication standards. The key system parameters are detailed in Table~\ref{tab1}. Utilizing the dataset described earlier in our simulations, we proceed to present our findings.
	
	\begin{table}[t]
		\caption{System parameters}
		\begin{center}
			\begin{tabular}{|r|c|l}
				\hline
				\hline
				\textbf{Parameter} 				&  \textbf{Value}	\\	\hline
				\hline
				Area of interest 				&  	5$\times$5 m$^2$  	\\      
				Height  of access node  					&	5 m		\\
				\hline
				Channel model  					&	mmAgic InH	\\
				Antenna model  					&	3GPP 		\\
				Traffic model  					&	Full buffer \\    			   		 
				Antenna model					& 3GPP antenna array\cite{Ant_3gpp} \\
				Access node antenna arrays		& 64 $\times$ 64 \\
				UE antenna arrays		& var. \\
				Beam management  				& Hierarchical codebook-based \\  
				Beacon interval  				& 100 ms \\  
				Sector-level sweep duration   				& 16 $\mu$s \\  
				Beam refinement duration   				& 1 $\mu$s \\  \hline						
				Carrier frequency 				&	60 GHz  \\  
				Carrier bandwidth				&	2.16 GHz  								\\  
				Receiver sensitivity 			&	-78 dBm		\\  
				Transmit power 					&	 0 dBm		\\  
				Noise figure					& 9 dB			\\  
				Power spectral density of noise	& 174 dBm/Hz	\\ 
				Maximum EIRP					& 40 dBm	\\ 
				Minimum SNR threshold			& -10.3 dB (MCS 0)	\cite{chen2017design}		\\  
				\hline\hline
			\end{tabular}
			\label{tab1}
		\end{center}
	\end{table}
	
	\subsubsection{Impact of Lateral Movements vs. Angular Movements}	
\begin{figure}[t]
		\centering 
		\includegraphics[width=0.9\linewidth]{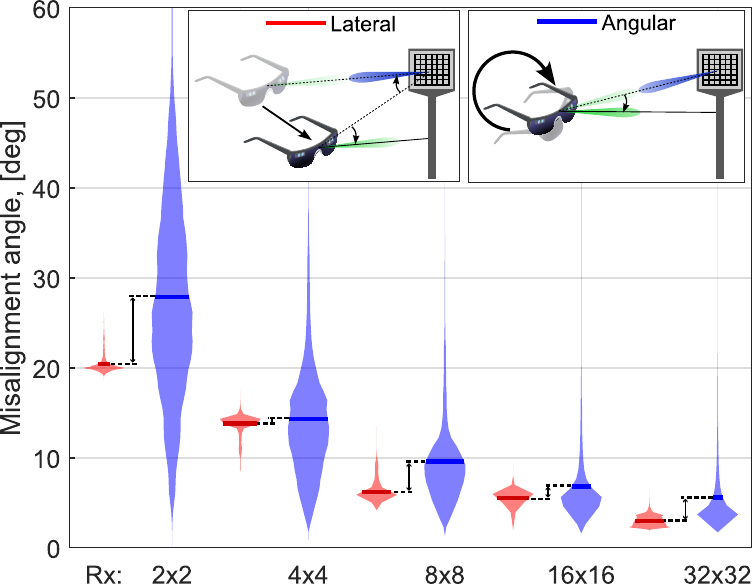}
		\caption{Misalignment due to angular and lateral movement.}
		\label{fig:result_1}
	\end{figure}

	To assess the impact of mobility on highly directional beams, we focus on beam misalignment, which is the difference between intended and actual beam directions. In Fig. \ref{fig:result_1}, we present cumulative distributions of beam misalignment for all 40 participants,  separately angular and lateral movements for different antenna array sizes. To analyze the effects of lateral movement, we maintain constant orientation angles and focus solely on the x, y, and z displacements of the UE to quantify their impact on misalignment. Conversely, to assess the influence of angular movement, we assume a constant lateral position while examining changes in UE orientation.  Our study investigates beam misalignment for varying numbers of antenna elements. In codebook-based beamforming, the number of beams aimed toward specific directions depend on the antenna array size. During each Beacon Interval (BI), the beam with the strongest signal is selected for the remainder of the BI, which may not be perfectly aligned. Smaller antenna arrays form wider beams, resulting in increased mean misalignment angles as compared to larger antenna arrays.
	
	Our results highlight the more pronounced influence of angular movement on beam misalignment. Furthermore, we observe that broader beams formed with fewer antenna elements exhibit higher misalignment degrees. After the the Sector Level Sweep~(SLS) and the Beam Refinement Protocol~(BRP), a new beam is chosen for each BI by prioritizing the highest directivity gain. Importantly, this selected beam is not necessarily perfectly aligned but is chosen for the highest gain even when mostly misaligned. Wider beams demonstrate greater tolerance for misalignment, which allows them to maintain sufficient gain even under considerable misalignment. In contrast, narrower beams have lower tolerance for misalignment, thereby requiring a new beam selection for even \mbox{minor misalignments.}
	
	The size of the antenna array is closely linked to the specific type of immersive device being used. Smaller arrays are a practical choice for AR glasses given their limited available space, while larger arrays are better suited for VR goggles with more surface area. Additionally, continuous technological advancements (such as smaller element spacing and on-lens antennas) should enable the reduction of space requirements for these arrays in the future, hence making them even more compact. It is important to emphasize that, for all UE antenna arrays, the effective isotropic radiated power (EIRP) has consistently remained below the maximum EIRP limit of 40 dBm, in compliance with FCC and ETSI regulations. These outcomes highlight the substantial effect of user orientation changes on XR communication system performance.

	\subsubsection{Impact of Angular Speed on Beam Misalignment}	
\begin{figure}[t]
		\centering 
		\includegraphics[width=0.99\linewidth]{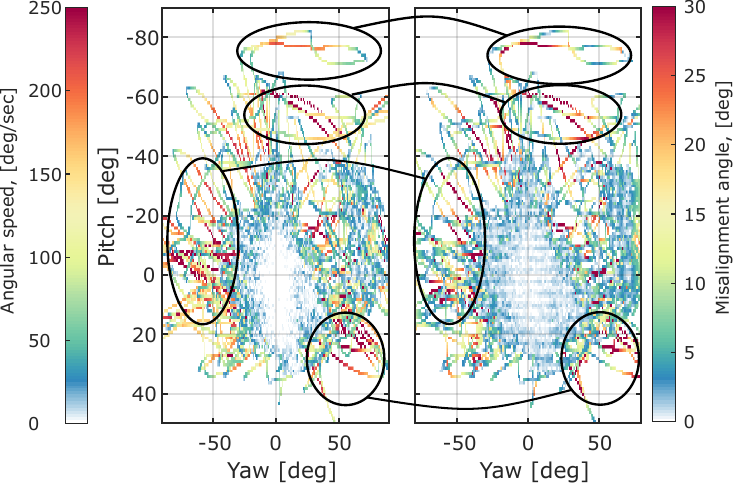} 
		\caption{Beam misalignment during VR gaming.}
		\label{fig:result_3}
	\end{figure}

	Fig.~\ref{fig:result_3} presents side-by-side comparisons of angular speeds and misalignment angles concerning the headset orientation for all participants. Notably, the central point of interest in VR gaming occurs when both pitch and yaw angles are close to 0. Our observations reveal that the angular speed at this central focus point is relatively slow, and as the user's gaze moves away from the center, the angular speed tends to increase. This phenomenon has potential implications for link utilization. The speed at which the user's head moves directly affects the frame size in XR applications, particularly during higher small-scale mobility characterized by the rapid angular speed of user \mbox{head rotation \cite{kim2020motion}.}
	
	The gradual shift in gaze around the central focus point can be due to the user's focus changing more gradually when gazing towards the center, potentially due to compensatory eye movements \cite{sidenmark2019eye}. In contrast, challenges arise when higher angular speeds manifest as the user explores the digital environment or shifts focus to the peripheries, leading to increased data transmission requirements. This exposes the headset to the risks of capacity degradation and outages, particularly when higher data rates and reliability are demanded.
	
	Additionally, the right side of Fig.~\ref{fig:result_3} displays the beam misalignment corresponding to the angular speed on the left side. The results reveal a clear relationship, indicating that higher angular speeds proportionally correspond to greater angles of beam misalignment. Such misalignment can lead to significant reductions in SNR hence directly impacting the QoE during immersive VR sessions by introducing buffering, lower resolution, and increased motion-to-photon latency. These findings highlight the importance of monitoring and accounting for user angular speed to optimize the wireless connection performance in VR scenarios, especially for delivering seamless and immersive experiences. This highlights the need for responsive beam management strategies, where beams can swiftly switch to compensate for misalignment.

	\subsubsection{Small-Scale Mobility Challenges in XR Communication}	
	To analyze the effects of angular speed on system performance, we employ synthetic user mobility traces, which capture angular speeds in a controlled manner. Synthetic data enable testing wireless performance across a wide range of parameters range, giving us control over the angular speed and facilitating the calculation of the mean performance values.
	
\begin{figure}[t]
		\centering 
		\includegraphics[width=0.95\linewidth]{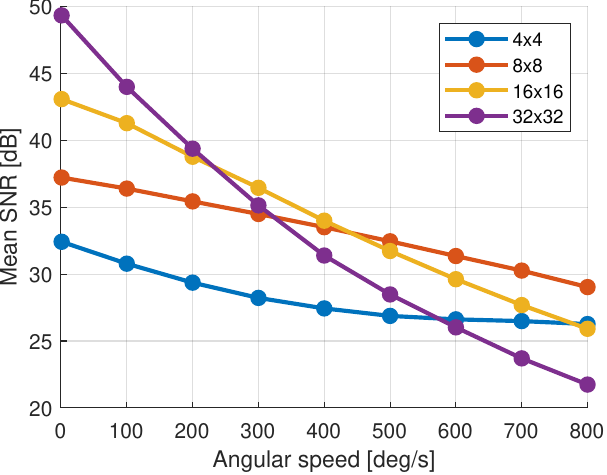} 
		\caption{SNR w.r.t. angular speed.}
		\label{fig:result_4}
	\end{figure}

	As illustrated in Fig.~\ref{fig:result_4}, it is evident that narrower beams perform better at lower angular speeds in terms of SNR. However, as the angular speed increases, the performance of the system employing narrower beams starts to deteriorate, leading to a decline in the mean SNR. This decrease can be attributed to the growing need for more frequent beam-switching at higher speeds, which causes substantial SNR drops during these transitions and potentially leads to connection disruptions or outages. Conversely, smaller antenna arrays forming wider beams offer broader coverage per beam, thereby reducing the necessity for frequent beam switches and associated performance decline. The wider beams benefit from better mean SNR at higher angular speeds, which suggests their suitability for scenarios involving faster user movement, while employing narrower beams during slower \mbox{user movement.}
	
\begin{figure}[t]
		\vskip\baselineskip
		\centering 
		\includegraphics[width=0.95\linewidth]{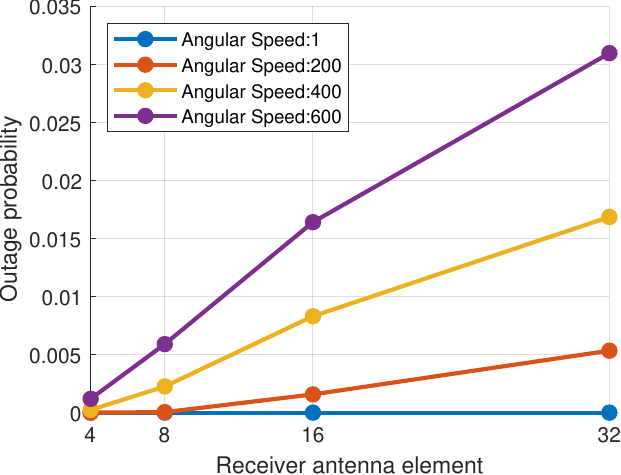} 
		\caption{Outage w.r.t. antenna elements.}
		\label{fig:result_5}
	\end{figure}

	In Fig.~\ref{fig:result_5}, we present the outage probability for varying numbers of antenna elements at different angular speeds. Outage probability, which is a crucial parameter for evaluating wireless system performance, represents the proportion of time when the SNR falls below a minimum threshold, hence leading to connection disruptions. Influenced by factors such as propagation conditions, receiver sensitivity, transmit power, and antenna arrays at both ends, outage probability increases with both higher numbers of antenna elements and angular speeds. This rise can be attributed to the growing frequency of beam switch events at higher speeds, where SNR drops drastically during transitions, by leading to outages and temporary connection losses. Additionally, transitioning from line-of-sight to non-line-of-sight beams as the user head turns away from the access node may trigger outages, which are more likely at higher speeds. For XR applications, this can cause frame judder, hence leading to double images, ghosting, and blurring, thereby disrupting the immersive experience.
	
	Overall, our findings from the simulations highlight the significant impact of user small-scale mobility, thereby particularly the angular speed, on mmWave link performance between the UE and the access node in XR applications. Wider beams perform better during faster user movement, while narrower beams yield better performance during slower user movement. This necessitates dynamic beamwidth adjustments according to the user movement. In summary, our findings emphasize the importance of optimizing beam management strategies and introducing redundancy in wireless systems to accommodate user mobility, ensure reliability, and enhance the performance of directional connections in immersive XR applications.
	
	
	\section{Key Outcomes and Conclusions}
	Due to the inherent nature of human behavior during an XR session, the UE experiences a significant degree of small-scale mobility. This fact, coupled with the stringent communication requirements of XR applications that demand highly directional communication, poses a formidable challenge in consistently delivering a resilient connection without outages. As XR services become increasingly prevalent, it is of significant importance for network architectures and protocols to evolve to meet the XR wireless connectivity requirements by providing reliable connectivity.
	
	In this article, we investigate the impact of small-scale user mobility on wireless XR communications, with a particular focus on its significance in the context of highly directional transmission. We provide a comprehensive overview of the small-scale mobility phenomenon during XR device usage and its influence on wireless communication using highly directional beams. Additionally, we discuss innovative approaches to address the challenges posed by small-scale mobility in XR applications. To deepen our understanding of small-scale mobility during consumer VR gaming, we collect a dataset encompassing the movements of 40 participants.
	
	Utilizing this dataset, we further conduct extensive simulations to study the effects of small-scale mobility on highly directional communication and analyze the output to identify the aspects of small-scale mobility that impact performance. Our dataset and results provide valuable insights for researchers and industry practitioners aiming to enhance wireless communication for XR applications. The dataset can serve as a benchmark for evaluating and comparing different mobility models and communication protocols to develop new beam \mbox{management mechanisms.}
	
	The key learning from our study is the notable influence of angular movements on beam misalignment as compared to lateral movements. We further investigate the relationship between angular speed and beam misalignment during VR gaming, by revealing that higher speeds lead to greater misalignment angles and potential SNR reductions, ultimately causing outages. Therefore, a careful consideration of the trade-off between directivity gain and link reliability is essential for the development of \mbox{next-generation XR systems.}

	\bibliographystyle{IEEEtran}
	\bibliography{refs_mag_final}
	
	\section*{Acknowledgements}
	This work was supported by the European Union’s Horizon 2020 Research and Innovation Programme under the Marie Skłodowska-Curie grant agreement No. 813278 (A-WEAR; http://www.a-wear.eu/). This work was also supported by the Research Council of Finland (Projects RADIANT, \mbox{ECO-NEWS}, and SOLID).

	\section*{Biography}
	\noindent{\bf{Asad Ali} }(asad.ali@tuni.fi) is a joint Ph.D. student at Tampere University, Finland, and Brno University of Technology (BUT), Czech Republic. He holds a B.Sc. in Electrical Engineering from COMSATS University, Pakistan, and an M.Sc. in Electrical Engineering from the University of Rostock, Germany. His research focuses on 5G/6G wireless networks and extended reality.
	
	\noindent{\bf{Olga Galinina}} (olga.galinina@tuni.fi) is a Senior Research Fellow at Tampere University, Finland. She earned her B.Sc. and M.Sc. degrees from Peter the Great St. Petersburg Polytechnic University and her Ph.D. from Tampere University of Technology in 2015. Her research interests include applied mathematical modeling and analysis of wireless networks and statistical machine learning.
	
	\noindent{\bf{Jiri Hosek}} (hosek@vut.cz) received his M.S. and Ph.D. in electrical engineering from BUT in 2007 and 2011, respectively. He is currently an associate professor (2016), deputy vice-head of R$\&$D and International Relations in the 	Department of Telecommunications (2018), and coordinates the WISLAB research group (since 2012) at the Department of Telecommunications, BUT. He has (co-)authored 130+ research works in networking, wireless communications, quality of service, quality of experience, and IoT applications.
	
	\noindent{\bf{Sergey Andreev}} (sergey.andreev@tuni.fi) is a Professor of wireless communications and an Academy Research Fellow at Tampere University, Finland. He is also a Research Specialist with Brno University of Technology, Brno, Czech Republic. He received his Ph.D. (2012) from TUT as well as his Specialist (2006), Cand.Sc. (2009), and Dr.Habil. (2019) degrees from SUAI. Sergey (co-)authored more than 300 published research works on intelligent IoT, mobile communications, and heterogeneous networking.
	
\end{document}